\newcommand{\amd}{{x86-64}}

\newcommand{\href}{{\tt h264ref}}
\newcommand{\bzip}{{\tt 401.bzip2}}

\newcommand{\hmmer}{{\tt 456.hmmer}}

\documentclass[10pt, conference, compsocconf]{IEEEtran}
\usepackage{mathptmx} 

\usepackage[normalem]{ulem}
\usepackage{color}
\usepackage{graphics}  
\usepackage{tablefootnote}
\usepackage{listings} 
\usepackage{fancyhdr} 

\usepackage{subcaption}
\usepackage{graphicx} 
\usepackage{adjustbox}
\usepackage{multirow}

\usepackage[hyphens]{url}
\usepackage{hyperref}

\newcommand{\wunits}[2]{\mbox{#1\,#2}}

\newcommand{\by}[2]{\mbox{#1$\times$#2}}

\title{\Large{\bf The Renewed Case for the Reduced Instruction Set Computer: \\
Avoiding ISA Bloat with Macro-Op Fusion for RISC-V}}

\author{\IEEEauthorblockN{Christopher Celio, Palmer Dabbelt, David Patterson, Krste Asanovi{\'{c}}}
\IEEEauthorblockA{Department of Electrical Engineering and Computer
  Sciences, University of California, Berkeley\\
celio@eecs.berkeley.edu}
}

\usepackage{etoolbox}
\apptocmd{\thebibliography}{\setlength{\itemsep}{8pt}}{}{}

\begin{document}

\date{}
\maketitle

\thispagestyle{empty}
\pagestyle{plain}

\begin{abstract}


This report makes the case that a well-designed Reduced Instruction 
Set Computer (RISC) can match, and even exceed, the performance 
and code density of existing commercial Complex Instruction Set 
Computers (CISC) while maintaining the simplicity and cost-effectiveness 
that underpins the original RISC goals~\cite{patterson1980case}.



We begin by comparing the dynamic instruction counts and dynamic
instruction bytes fetched for the popular proprietary ARMv7, ARMv8,
IA-32, and \amd\ Instruction Set Architectures (ISAs) against the free
and open RISC-V RV64G and RV64GC ISAs when running the SPEC CINT2006
benchmark suite.  RISC-V was designed as a very small ISA to support a
wide range of implementations, and has a less mature compiler
toolchain.  However, we observe that on SPEC CINT2006 RV64G executes 
on average 16\% more instructions than x86-64, 3\% more instructions than 
IA-32, 9\% more instructions than ARMv8, but 4\% fewer instructions than ARMv7.

CISC x86 implementations break up complex instructions
into smaller internal RISC-like {\em micro-ops}, and the RV64G instruction count is
within 2\% of the \amd\ retired micro-op count.  RV64GC, the
compressed variant of RV64G, is the densest ISA studied, fetching 8\%
fewer dynamic instruction bytes than \amd.  We observed that much of
the increased RISC-V instruction count is due to a small set of common
multi-instruction idioms.  

Exploiting this fact, the RV64G and RV64GC {\em effective
instruction} count can be reduced by 5.4\% on average by leveraging
{\em macro-op fusion}.  Combining the compressed RISC-V ISA extension with macro-op
fusion provides both the densest ISA and the fewest dynamic
operations retired per program, reducing the motivation to add more
instructions to the ISA.  This approach retains a single simple ISA
suitable for both low-end and high-end implementations, where high-end
implementations can boost performance through microarchitectural
techniques.

{\em Compiler tool chains are a continual work-in-progress, and the
  results shown are a snapshot of the state as of July 2016 and are
  subject to change.}

\end{abstract}


\section{Introduction}\label{sec:introduction}

The Instruction Set Architecture (ISA) specifies the set of
instructions that a processor must understand and the expected
effects of each instruction.  One of the goals of the
RISC-V project was to produce an ISA suitable for a wide range of
implementations from tiny microcontrollers to the largest
supercomputers~\cite{Waterman:EECS-2014-54}. 
Hence, RISC-V was designed with a
much smaller number of simple standard instructions compared to other
popular ISAs, including other RISC-inspired ISAs.  A simple ISA is
clearly a benefit for a small resource-constrained microcontroller,
but how much performance is lost for high-performance implementations
by not supporting the numerous instruction variants provided by
popular proprietary ISAs?

A casual observer might argue that a processor's performance increases 
when it executes fewer instructions for a given program, but in reality, 
the performance is more accurately described by the Iron Law of 
Performance~\cite{hennessy2011computer}:

\begin{quote}
$\frac{seconds}{program} = \frac{cycles}{instruction} * \frac{seconds}{cycle} *\frac{instructions}{program}$
\end{quote}

The ISA is just an abstract boundary; behind the scenes the processor may
choose to implement instructions in any number of ways that trade off
$\frac{cycles}{instruction}$, or {\em CPI}, and  $\frac{seconds}{cycle}$, or
{\em frequency}.  

For example, a fairly powerful x86 instruction is the {\em repeat move}
instruction ({\tt rep movs}), which copies $C$ bytes of data from one
memory location to another:

{\footnotesize
\begin{verbatim}
  // psuedo-code for a `repeat move' instruction
  for (i=0; i < C; i++)
     d[i] = s[i];
\end{verbatim}
}

Implementations of the x86 ISA break up the {\em repeat move}
instruction into smaller operations, or {\em micro-ops}, that
individually perform the required operations of loading the data from
the old location, storing the data to the new location, incrementing
the address pointers, and checking to see if the end condition has
been met.  Therefore, a raw comparison of instruction counts may hide
a significant amount of work and complexity to execute a particular
benchmark.


In contrast to the process of generating many micro-ops from a single ISA instruction, 
several commercial microprocessors perform
{\em macro-op fusion}, where several ISA instructions are fused in the
decode stage and handled as one internal operation.  As an example,
compare-and-branch is a very commonly executed idiom, and the RISC-V
ISA includes a full register-register magnitude comparison in its
branch instructions.  However, both ARM and x86 typically require two ISA
instructions to specify a compare-and-branch. The first instruction
performs the {\em comparison} and sets a condition code, and the
second instruction performs the {\em jump-on-condition-code}.  While
it would seem that ARM and x86 would have a penalty of one additional instruction
on nearly every loop compared to RISC-V, the reality is more
complicated.  Both ARM and Intel employ the technique of macro-op
fusion, in which the processor front-end detects these
two-instruction compare-and-branch sequences in the instruction stream
and ``fuses'' them together into a single {\em macro-op}, which can then be
handled as a single compare-and-branch instruction by the processor
back-end to reduce the effective dynamic instruction count.\footnote{The reality can be even more complicated. Depending on the micro-architecture, the front-end may fuse the two instructions together to save decode, allocation, and commit bandwidth, but break them apart in the execution pipeline for critical path or complexity reasons~\cite{gochman2003intel}.}

Macro-op fusion is a very powerful technique to lower the effective
instruction count.  One of the main contributions of this report is to
show that macro-op fusion, in combination with the existing compressed
instruction set extensions for RISC-V, can provide the effect of a
richer instruction set for RISC-V without requiring any ISA
extensions, thus enabling support for both low-end implementations and
high-end implementations from a single simple common code base.  The
resulting ISA design can provide both a low number of effective
instructions executed and a low number of dynamic instruction bytes
fetched.

\begin{table*}[b!htbp]
\caption{Compiler options for {\tt gcc 5.3}.}
\begin{center}
\begin{tabular}{|c|c|l|}
\hline
ISA & compiler & flags \\
\hline
RV64G & riscv64-unknown-gnu-linux-g++ & -O3 -static  \\
RV64GC & riscv64-unknown-gnu-linux-g++ & -O3 -mrvc -mno-save-restore -static  \\
IA-32 & g++-5 & -O3 -m32 -march=ivybridge -mtune=native -static\\
x86-64 & g++-5 & -O3 -march=ivybridge -mtune=native -static\\
ARMv7ve & g++ & -O3 -march=armv7ve -mtune=cortex-a15 -static\\
ARMv8-a & g++-5 & -O3 -march=armv8-a -mtune=cortex-a53 -static \\
& & -mfix-cortex-a53-835769 -mfix-cortex-a53-843419 \\
\hline
\end{tabular}
\end{center}
\label{table:cflags}
\end{table*}%

\section{Methodology}\label{sec:methodology}

In this section, we describe the benchmark suite and methodology used
to obtain dynamic instruction counts, dynamic instruction bytes, and effective instructions
executed for the ISAs under consideration.

\subsection{SPEC CINT2006}

We used the SPEC CINT2006 benchmark suite~\cite{henning2006spec} for
comparing the different ISAs. SPECInt2006 is composed of 35 different
workloads across 12 different benchmarks with a focus on desktop and
workstation-class applications such as compilation, simulation,
decoding, and artificial intelligence.  These applications are largely
CPU-intensive with working sets of tens of megabytes and a required
total memory usage of less than \wunits{2}{GB}.

\subsection{GCC Compiler}

We used GCC for all targets as it is widely used and the only
compiler available for all systems.  Vendor-specific compilers will
surely provide different results, but we did not analyze them here.
All benchmarks were compiled using the latest {\tt GNU gcc 5.3} with
the parameters shown in Table \ref{table:cflags}. The {\tt
  400.perlbench} benchmark requires specifying {\tt -std=gnu98} to
compile under {\tt gcc 5.3}.  We used the {\tt Speckle} suite to
compile and execute SPECInt using {\tt reference} inputs to completion
\cite{speckle}.
The benchmarks were compiled {\em statically} to make it easier to analyze the binaries.
Unless otherwise specified, data was collected using the {\tt perf} utility~\cite{perf} while running the benchmarks on native hardware.

\subsection{RISC-V RV64}

The RISC-V ISA is a free and open ISA produced by the University of
California, Berkeley and first released in 2010~\cite{riscv}.  For
this report, we will use the standard RISC-V RV64G ISA variant, which
contains all ISA extensions for executing 64-bit ``general-purpose"
code~\cite{Waterman:EECS-2014-54}.  We will also explore the ``C" Standard Extension
for Compressed Instructions (RVC).  All instructions in RV64G are 4-bytes in
size, however, the C extension adds 2-byte forms of the most common
instructions. The resulting RV64GC ISA is very dense, both statically
and dynamically~\cite{Waterman:EECS-2016-1}.

We cross-compiled RV64G and RV64GC benchmarks using the compiler
settings shown in Table~\ref{table:cflags}. The RV64GC benchmarks were
built using a compressed glibc library.  The benchmarks were then
executed using the {\tt spike} ISA simulator running on top of {\tt Linux} 
version 3.14, which was compiled against version 1.7 of the RISC-V privileged ISA.
A side-channel process grabbed the retired instruction
count at the beginning and end of each workload.
We did not analyze RV32G, as there does not yet exist an RV32 port of
the {\tt Linux} operating system.

For the {\tt 483.xalancbmk} benchmark, 34\% of the RISC-V instruction
count is taken up by an OS kernel spin-loop waiting on the
test-harness I/O. These instructions are an artifact of our testing
infrastructure and were removed from any further analysis.

\subsection{ARMv7}

The 32-bit ARMv7 benchmarks were compiled and executed on an Samsung
Exynos 5250 (Cortex A-15). The {\tt march=native} flag resolves to the
ARMv7ve ISA and the {\tt mtune=native} flag resolves to the {\tt cortex-a15}
processor.

\subsection{ARMv8}

The 64-bit ARMv8 benchmarks were compiled and executed on a Snapdragon
410c (Cortex A-53). The {\tt march} flag was set to the ARMv8-a ISA and
the {\tt mtune} flag was set to the {\tt cortex-a53} processor. The errata
flags for {\tt -mfix-cortex-a53-835769} and {\tt
  -mfix-cortex-a53-843419} are set.  The \wunits{1}{GB} of RAM on the
410c board is not sufficient to run some of the workloads from {\tt
  401.bzip2}, {\tt 403.gcc}, and {\tt 429.mcf}.  To manage this issue,
we used a swapfile to provide access to a larger pool of memory and
only measured user-level instruction counts for the problematic
workloads.

\subsection{IA-32}

The architecture targeted is the {\tt i686} architecture and was
compiled and executed on an Intel Xeon E5-2667v2 (Ivy Bridge).

\subsection{\amd}

The \amd\ benchmarks were compiled and executed on an Intel Xeon
E5-2667v2 (Ivy Bridge).  The {\tt march} flag resolves to the
{\tt ivybridge} ISA.

\subsection{Instruction Count Histogram Collection}

Histograms of the instruction counts for RV64G, RV64GC, and \amd\ were
collected allowing us to more easily compare the hot loops across
ISAs.  We were also able to compute the dynamic instruction bytes by
cross-referencing the histogram data with the static objdump data.
\amd\ histograms were collected by writing a histogram-building tool for
the Intel Pin dynamic binary translation tool~\cite{pintool}.
Histograms for RV64G and RV64GC were collected using an existing
histogram tool built into the RISC-V {\tt spike} ISA simulator.

\subsection{SIMD ISA Extensions}

Although a vector extension is planned for RISC-V, there is no
existing vector facility.  To compare against the scalar RV64G ISA, we
verified that the ARM and x86 code were compiled in a manner that
generally avoided generating any SIMD or vector instructions for the
SPECInt2006 benchmarks.  An analysis of the \amd\ histograms showed that, with
the exception of the {\tt memset} routine in {\tt 403.gcc} and a {\tt strcmp} 
routine in {\tt 471.omnetpp}, no SSE instructions were generated that 
appeared in the 80\% most executed instructions.

To further reinforce this conclusion, we built a {\tt gcc} and {\tt
  glibc} \amd\ toolchain that explicitly forbade MMX and AVX
extensions. Vectorization analysis was also disabled. The resulting
instruction counts for SPECInt2006 were virtually unchanged.

Although the MMX and AVX extensions may be disabled in {\tt gcc}, it
is not possible to disable SSE instruction generation as it is a
mandatory part of the \amd\ floating point ABI.  However, we note that
the only significant usage of SSE instructions were 128-bit SSE stores found in
the {\tt memset} routine in {\tt 403.gcc} ($\approx$20\%) and a very small usage ($<$2\%) of packed SIMD found in {\tt strcmp} in {\tt 471.omnetpp}.




\section{Results}\label{sec:results}
 
All comparisons between ISAs in this report are based on the geometric
mean across the 12 SPECInt2006 benchmarks.

\begin{figure*}
        \centering
{\includegraphics[scale=1.0]{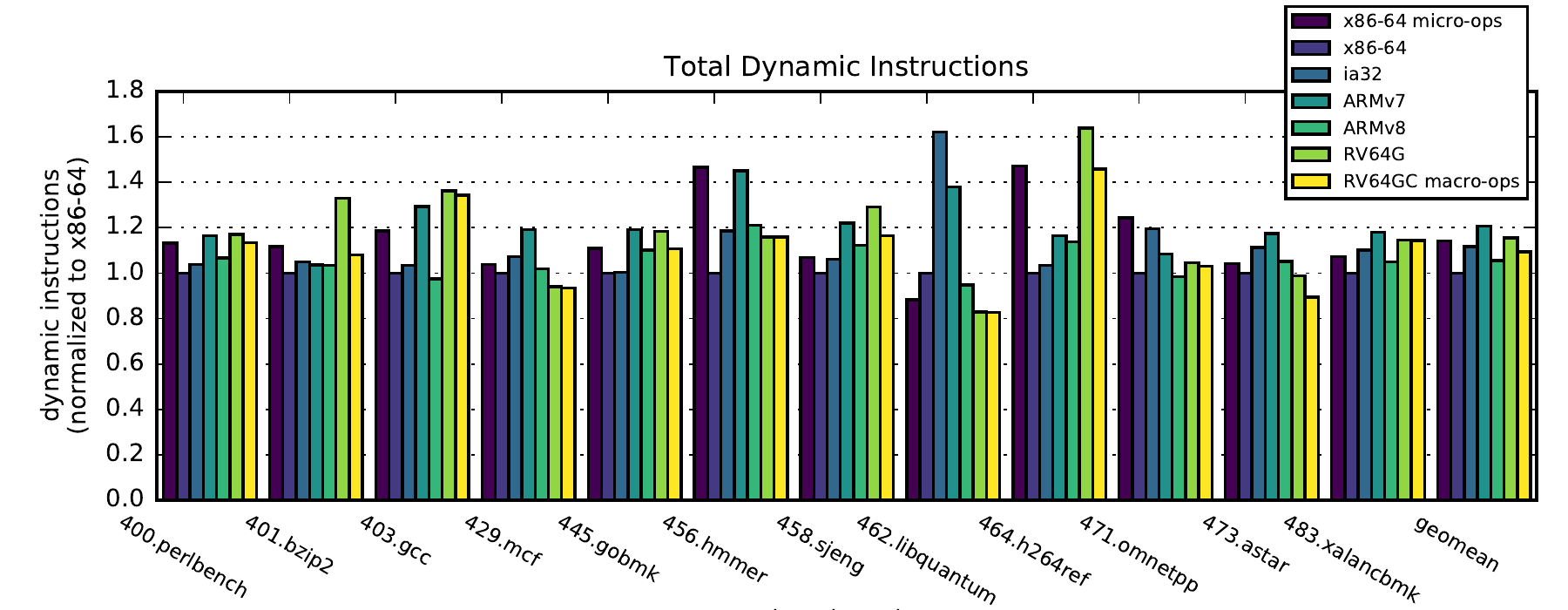}}
        \caption{The total dynamic instruction count is shown for each of the
            ISAs, normalized to the \amd\ instruction count. The \amd\ retired
            micro-op count is also shown to provide a comparison between
            \amd\ instructions and the actual operations required to
            execute said instructions.
            By leveraging macro-op fusion (in which some common multi-instruction
            idioms are combined into a single operation), the ``effective''
            instruction count for RV64GC can be reduced by 5.4\%.
        }
        \label{fig:instcounts}
{\includegraphics[scale=1.0]{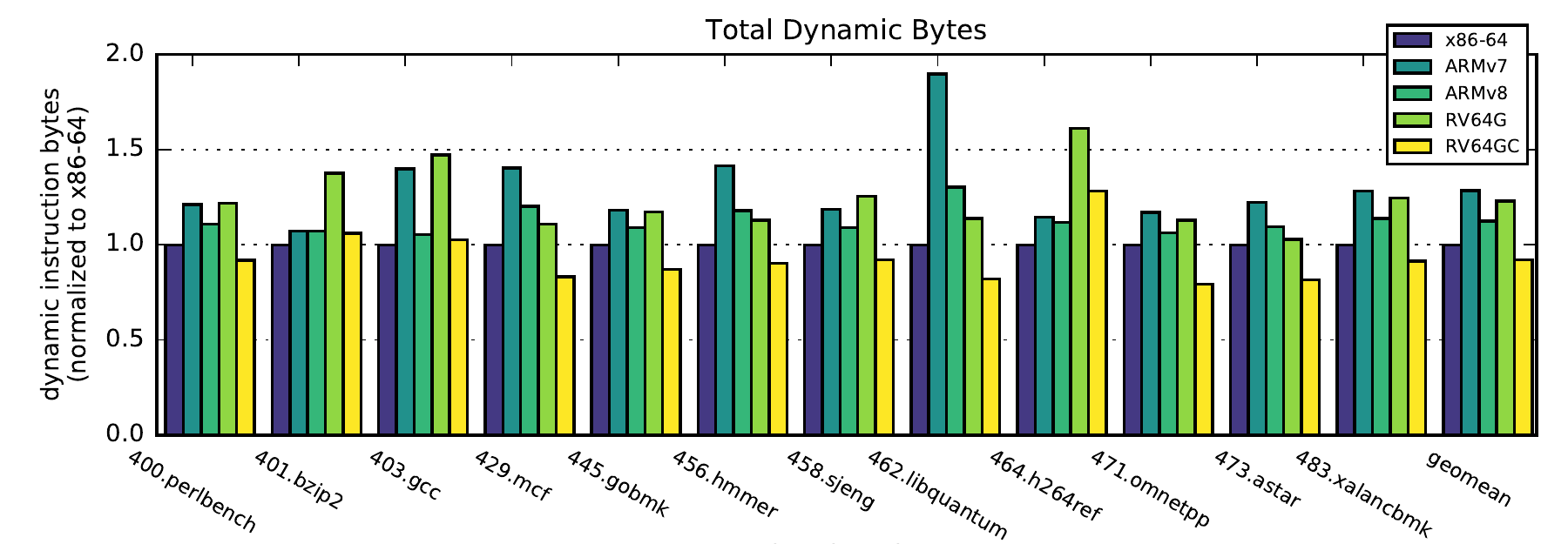}}
        \caption{Total dynamic bytes normalized to \amd.  RV64G, ARMv7, and
            ARMv8 use fixed 4 byte instructions. \amd\ is a variable-length ISA
            and for SPECInt averages 3.71 bytes / instruction. RV64GC uses
            two byte forms of the most common instructions allowing it to
            average 3.00 bytes / instruction.}
        \label{fig:dynamicbytes}
\end{figure*}

\begin{figure*}
        \centering
{\includegraphics[scale=1.0]{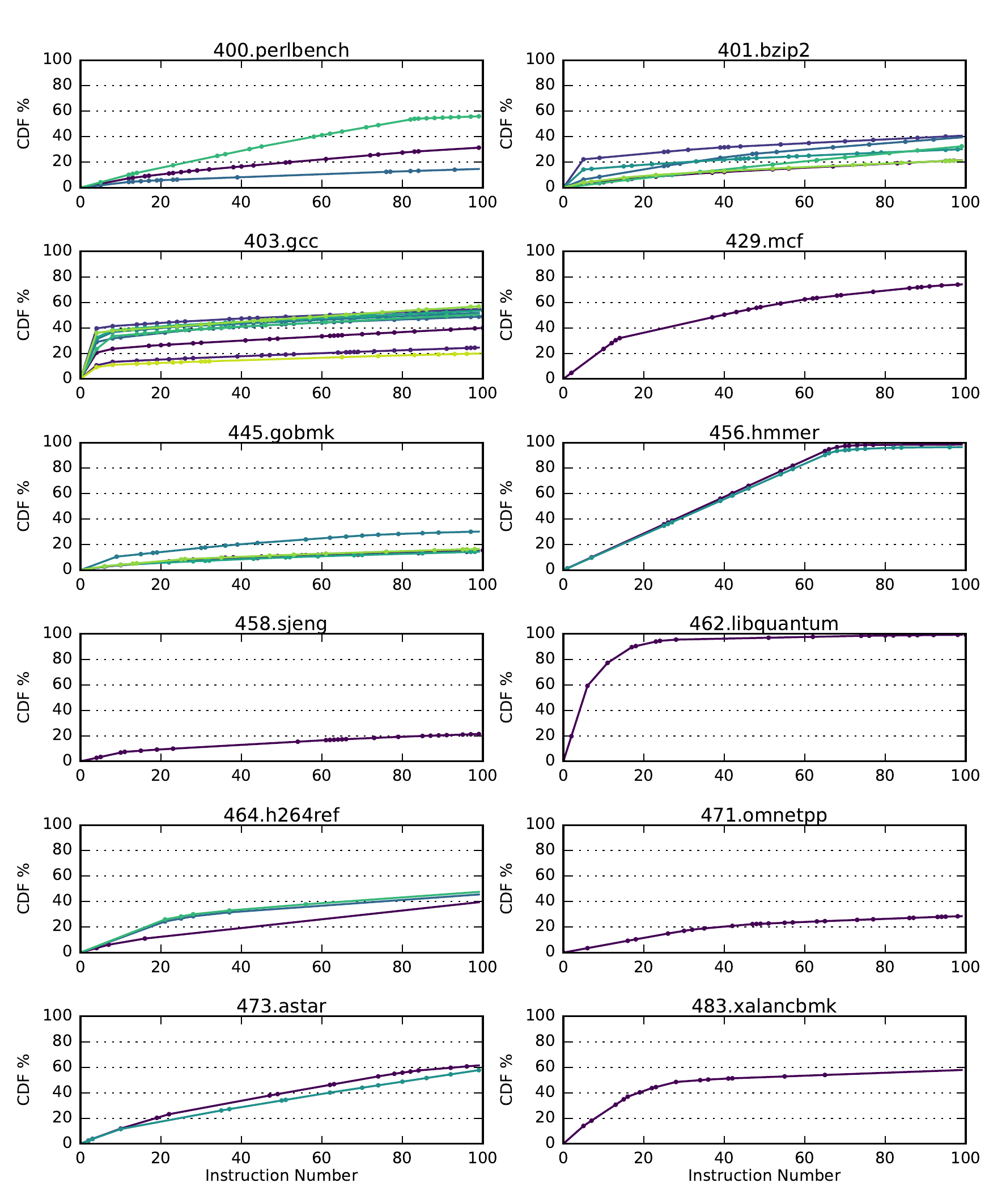}}
        \caption{Cumulative distribution function for the 100 most frequent
            RISC-V instructions of each of the 35 SPECInt workloads.  Each line
            corresponds to one of the 35 SPECInt workloads. Some SPECInt
            benchmarks only have one workload.  A (*) marker denotes the
            start of a new contiguous instruction sequence (that ends with
            a taken branch).}
        \label{fig:cdfs}
\end{figure*}

\subsection{Instruction Counts}

\setlength{\tabcolsep}{0.1em} 
\begin{table}[htp]
\caption{Total dynamic instructions normalized to \amd.}
\begin{center}
\begin{tabular}{|c|ccccccc|}
\hline
benchmark       & \multicolumn{1}{|c|}{\amd} & \multicolumn{1}{|c|}{\amd} & \multicolumn{1}{|c|}{IA-32} & \multicolumn{1}{|c|}{ARMv7} & \multicolumn{1}{|c|}{ARMv8} & \multicolumn{1}{|c|}{RV64G} & \multicolumn{1}{|c|}{RV64GC+}\\
                & \multicolumn{1}{|c|}{micro-ops} & \multicolumn{1}{|c|}{}  & \multicolumn{1}{|c|}{}      & \multicolumn{1}{|c|}{}      &    \multicolumn{1}{|c|}{}   & \multicolumn{1}{|c|}{}      & \multicolumn{1}{|c|}{fusion} \\
\hline
400.perlbench   & 1.13  & 1.00  & 1.04  & 1.16  & 1.07  & 1.17  & 1.14  \\
401.bzip2       & 1.12  & 1.00  & 1.05  & 1.04  & 1.03  & 1.33  & 1.08  \\
403.gcc         & 1.19  & 1.00  & 1.03  & 1.29  & 0.97  & 1.36  & 1.34  \\
429.mcf         & 1.04  & 1.00  & 1.07  & 1.19  & 1.02  & 0.94  & 0.93  \\
445.gobmk       & 1.11  & 1.00  & 1.00  & 1.19  & 1.10  & 1.18  & 1.11  \\
456.hmmer       & 1.47  & 1.00  & 1.19  & 1.45  & 1.21  & 1.16  & 1.16  \\
458.sjeng       & 1.07  & 1.00  & 1.06  & 1.22  & 1.12  & 1.29  & 1.16  \\
462.libquantum  & 0.88  & 1.00  & 1.62  & 1.38  & 0.95  & 0.83  & 0.83  \\
464.h264ref     & 1.47  & 1.00  & 1.03  & 1.17  & 1.14  & 1.64  & 1.46  \\
471.omnetpp     & 1.24  & 1.00  & 1.20  & 1.08  & 0.98  & 1.05  & 1.03  \\
473.astar       & 1.04  & 1.00  & 1.11  & 1.17  & 1.05  & 0.99  & 0.89  \\
483.xalancbmk   & 1.07  & 1.00  & 1.10  & 1.18  & 1.05  & 1.15  & 1.14  \\
\hline
geomean         & 1.14  & 1.00  & 1.12  & 1.21  & 1.06  & 1.16  & 1.09  \\
\hline
\end{tabular}
\end{center}
\label{table:instcounts}
\end{table}

As shown in Figure 
\ref{fig:instcounts} (and Table~\ref{table:instcounts}), RV64G executes 16\% more instructions than \amd,
3\% more instructions than IA-32, 9\% more instructions than ARMv8, and
4\% fewer instructions than ARMv7.
The raw instruction counts can be found in Figure \ref{table:rawinstcounts}.

\subsection{Micro-op Counts}

The number of \amd\ {\em retired micro-ops} was also collected and is
reported in Figure 
\ref{fig:instcounts}.  On average, the Intel Ivy Bridge processor used in this study emitted 1.14 micro-ops per
\amd\ instruction, which puts the RV64G instruction count within 2\%
of the \amd\ {\em retired micro-op} count.

\subsection{Dynamic Instruction Bytes}

\setlength{\tabcolsep}{0.30em} 
\begin{table}[htp]
\caption{Total dynamic bytes normalized to \amd.}
\begin{center}
\begin{tabular}{|c|ccccc|}
\hline
benchmark & \multicolumn{1}{|c|}{\amd\ } & \multicolumn{1}{|c|}{ARMv7} & \multicolumn{1}{|c|}{ARMv8} & \multicolumn{1}{|c|}{RV64G} & \multicolumn{1}{|c|}{RV64GC} \\
\hline
400.perlbench   & 1.00  & 1.21  & 1.11  & 1.22  & 0.92  \\
401.bzip2       & 1.00  & 1.07  & 1.07  & 1.38  & 1.06  \\
403.gcc         & 1.00  & 1.40  & 1.05  & 1.47  & 1.03  \\
429.mcf         & 1.00  & 1.40  & 1.20  & 1.11  & 0.83  \\
445.gobmk       & 1.00  & 1.18  & 1.09  & 1.17  & 0.87  \\
456.hmmer       & 1.00  & 1.41  & 1.18  & 1.13  & 0.90  \\
458.sjeng       & 1.00  & 1.19  & 1.09  & 1.25  & 0.92  \\
462.libquantum  & 1.00  & 1.90  & 1.30  & 1.14  & 0.82  \\
464.h264ref     & 1.00  & 1.14  & 1.12  & 1.61  & 1.28  \\
471.omnetpp     & 1.00  & 1.17  & 1.06  & 1.13  & 0.79  \\
473.astar       & 1.00  & 1.22  & 1.10  & 1.03  & 0.82  \\
483.xalancbmk   & 1.00  & 1.28  & 1.14  & 1.24  & 0.91  \\
\hline
geomean         & 1.00  & 1.28  & 1.12  & 1.23  & 0.92  \\
\hline
\end{tabular}
\end{center}
\label{table:dynamicbytes}
\end{table}

The total dynamic instruction bytes fetched is reported in
Figure~\ref{fig:dynamicbytes} (and Table~\ref{table:dynamicbytes}).  RV64G, with its fixed 4-byte
instruction size, fetches 23\% more bytes per program than \amd.
Unexpectedly, \amd\ is not very dense, averaging 3.71 bytes per
instruction (with a standard deviation of 0.34 bytes).  Like RV64G,
both ARMv7 and ARMv8 use a fixed 4-byte instruction size.

Using the RISC-V ``C'' Compressed ISA extension, RV64GC fetches 8\%
fewer dynamic instruction bytes relative to \amd, with an average of
3.00 bytes per instruction.  There are only three benchmarks ({\tt
  401.bzip2}, {\tt 403.gcc}, {\tt 464.h264ref}) where RV64GC fetches
more dynamic bytes than \amd, and two of those three benchmarks make
heavy use of {\tt memset} and {\tt memcpy}.  RV64GC also fetches
considerably fewer bytes than either ARMv7 or ARMv8.

\section{Discussion}\label{sec:discussion}


We discuss briefly the three outliers where RISC-V performs poorly, as
well as general trends observed across all of the benchmarks for
RISC-V code.  A more detailed analysis of the individual benchmarks
can be found in the Appendix. 

\noindent{\bf 401.bzip2:} Array indexing is implemented using {\em
  unsigned int (32-bit)} variables.  This represents a case of poor
coding style, as the C code should have been written to use the
standard {\tt size\_t} type to allow portability to different address
widths.  Because RV64G lacks unsigned arithmetic operations on
sub-register-width types, and the RV64G ABI behavior is to sign-extend
all 32-bit values into signed 64-bit registers, a two-instruction
idiom is required to clear the upper 32-bits when compiler analysis
cannot guarantee that the high-order bits are not zero.

\noindent{\bf 403.gcc:} 30\% of the RISC-V instruction count is taken
up by a {\tt memset} loop. \amd\ utilizes a {\tt movdqa} instruction
(aligned double quad-word move, i.e., a 128-bit store) and a four-way
unrolled loop to move 64 bytes in 7 instructions versus RV64G's 4
instructions to move 16 bytes.

\noindent{\bf 464.h264ref:} 25\% of the RISC-V instruction count is
taken up by a {\tt memcpy} loop.  Those 21 RV64G instructions together
account for 1.1 trillion fetches, compared to a single \amd\ ``repeat
move'' instruction that is executed 450 billion times.

\noindent{\bf Remaining benchmarks}

Consistent themes of the remaining benchmarks are as follows:

\begin{itemize}
\item RISC-V's fused compare-and-branch instruction allows it to
  execute typical loops using one less instruction compared to the ARM
  and x86 ISAs, both of which separate out the comparison and the
  jump-on-condition into two distinct instructions.
\item Indexed loads are an extremely common idiom.  Although
  \amd\ and ARM implement indexed loads (register+register addressing
  mode) as a single instruction, RISC-V requires up to three
  instructions to emulate the same behavior.
\end{itemize}

In summary, when RISC-V is using fewer instructions relative to other
ISAs, the code likely contains a significant number of branches. When
RISC-V is using more instructions, it is often due to a significant
number of indexed memory operations, unsigned integer array indexing,
or library routines such as {\tt memset} or {\tt memcpy}.

We note that both {\tt memcpy} and {\tt memset} are ideal candidates
for vectorization, and that some of the other indexed memory
operations can be subsumed into vector memory load and store
instructions when the RISC-V vector extension becomes available.
However, in this report we focus on making improvements to a purely
scalar RISC-V implementation.

\section{A Devil's Argument: Add Indexed Loads to RISC-V?}

The indexed load is a common idiom for {\tt array[offset]}.  Given the
data discussed previously, it is tempting to ponder the addition of indexed
loads to RISC-V.
 
{\footnotesize
\begin{verbatim}
    // rd = array[offset]
    // where rs1 = &(array), rs2 = offset
    add rd, rs1, rs2
    ld  rd, 0(rd)
\end{verbatim}
}

A simple indexed load fulfills a number of requirements of a RISC
instruction:

\begin{itemize}
\item reads two source registers
\item writes one destination register
\item performs only one memory operation
\item fits into a 4-byte instruction
\item has the same side-effects as the existing load instruction
\end{itemize}

This is a common instruction in other ISAs.  For example, ARM calls
this a ``load with register offset'' and includes a small {\em shift}
to scale the offset register into a data-type aligned offset:
\footnote{Despite the claims that ARM is a RISC ISA (it's literally
  the `R' in their name, after all!), ARM's {\em load with register
    offset} (LDR) is just one example of how CISC-y ARM can be. The
  {\em LDR with pre/post-indexing} instruction can be masked off by a
  condition, it can perform up to two separate memory loads to two
  different registers, it can modify the base address source register,
  and it can throw exceptions. Better yet, LDR can write to the PC
  register in ARMv7 (and earlier) and thus turn the LDR into a
  (conditional) branch instruction that can even change the ISA mode!
  In other words, a single post-indexed LDR instruction using the
  stack pointer as the base address and writing to multiple registers,
  one of which is the PC, can be used to implement a {\em stack-pop
    and return from function call}.}

{\footnotesize
\begin{verbatim}
    // if (cond) Rt = mem[Rn +/- (Rm << shift)]
    LDR{type}{cond} Rt, [Rn +/- Rm {, shift}]
\end{verbatim}
}

ARM also includes post- and pre-indexed versions that increment the
base address register which requires an additional write port on the
register file.

The x86 ISA provides indexed loads that include both the scaling shift
and an immediate offset:

{\footnotesize
\begin{verbatim}
    // rsi = mem[rdx + rax*n + b]
    mov b(%rdx,%rax,n),%rsi
\end{verbatim}}

\section{The Angelic Response: Use Macro-op Fusion!}\label{sec:fusion}

While the indexed load is perhaps a compelling addition to a RISC ISA,
the same effect can be obtained using the RISC-V ``C'' Compressed
Extension (RVC) coupled with macro-op fusion.  {\bf Given the usage of
  RVC, the indexed load idiom in RISC-V becomes a \by{two}{two-byte}
  instruction sequence.}  This sequence can be fused in the processor
front-end to effect the same outcome as having added 4-byte indexed
loads to RISC-V proper.

There are other reasons to eschew indexed loads in the ISA. First, it
would be odd to not maintain symmetry by also adding an indexed store
instruction.\footnote{The Intel i860~\cite{i860architecture} took the
asymmetric approach of only adding register indexing to loads and
only supporting post-increment addressing for stores and floating-point memory operations.}  Indeed, the {\tt
gcc} compiler assumes that loads and stores utilize the same
addressing modes.  Unfortunately, while indexed loads can be quite
simple and cheap, indexed stores require a third register read port to
access the store data.  For RISC-V, indexed stores would be the first
and only three-operand integer instruction.

The rest of this section will explore macro-op fusion and measure the
potential reduction in ``effective'' instruction counts.

\subsection{Fusion Pair Candidates}

The following idioms are additional good candidates for macro-op
fusion.  Note that for macro-op fusion to take place, the first
instruction's destination register must be clobbered by the subsequent
instruction in the idiom such that only a single 
architectural register write is
observed. Also note that the RVC compressed ISA is not necessary to
utilize macro-op fusion: a pair of 4-byte instructions (or even a
2-byte and a 4-byte pair) can be fused with the same benefits.

\noindent{\bf Load Effective Address (LEA)}

The LEA idiom computes the effective address of a memory location and
places the address into a register. The typical use-case is an array
offset that is 1) shifted to a data-aligned offset and then 2) added
to the array's base address.

{\footnotesize
\begin{verbatim}
    // &(array[offset])
    slli rd, rs1, {1,2,3}
    add  rd, rd,  rs2
\end{verbatim}
}

\noindent{\bf Indexed Load}

The Indexed Load idiom loads data from an address computed by summing two registers.

{\footnotesize
\begin{verbatim}
    // rd = array[offset]
    add rd, rs1, rs2
    ld  rd, 0(rd)
\end{verbatim}
}

This pattern can be combined with the LEA idiom to form a single three-instruction fused indexed load:

\newpage
{\footnotesize
\begin{verbatim}
    // rd = array[offset]
    slli rd, rs1, {1,2,3}
    add  rd, rd, rs2
    ld   rd, 0(rd)
\end{verbatim}
}

\noindent{\bf Clear Upper Word}

The Clear Upper Word idiom zeros the upper 32-bits of a 64-bit
register. This often occurs when software is written using {\tt
  unsigned int} as an array index variable; the compiler must clear
the upper word to avoid potential overflow issues.\footnote{RISC-V
  matches the behavior of MIPS and Alpha. Registers hold signed
  values, but software must clear the high 32-bits when using an unsigned 32b
  to access an array. ARMv8 can perform such accesses in a single
  instruction, as it uses the register specifiers w0-w30 to access the
  bottom 32-bits of its 64-bit integer registers: (e.g., {\bf ldr w0, [x0,
      w1, uxtw 2]}).}

{\footnotesize
\begin{verbatim}
    // rd = rs1 & 0xffffffff
    slli rd, rs1, 32
    srli rd, rd,  32
\end{verbatim}
}

We also measure the occurrences of the Clear Upper Word idiom
followed by a small left shift by a few bits for aligning the register
to a particular data offset size, which appears as follows in assembly
code:

{\footnotesize
\begin{verbatim}
    slli rd, rs1, 32
    srli rd, rd,  {29,30,31,32}
\end{verbatim}
}

\noindent{\bf Load Immediate Idioms (LUI-based idioms)}

The {\em load upper immediate} (LUI) instruction is used to help construct immediate values that are larger than the typical 12 bits available to most RISC-V instructions. There are two particular idioms worth discussing. The first loads a 32-bit immediate into a register:

{\footnotesize
\begin{verbatim}
    // rd = imm[31:0]
    lui   rd, imm[31:12]
    addi  rd, rd, imm[11:0]
\end{verbatim}
}

Although the most common form is LUI/ADDI, it is perfectly reasonable to fuse any integer register-immediate instruction that follows a LUI instruction.

The second LUI-based idiom loads a value in memory statically addressed by a 32-bit immediate:

{\footnotesize
\begin{verbatim}
    // rd = *(imm[31:0])
    lui  rd, imm[31:12]
    ld   rd, imm[11:0](rd)
\end{verbatim}
}

Both of these LUI-based idioms are fairly trivial additions to any RISC pipeline.  However, we note that their appearance is SPECInt is less than 1\% and so we do not explore them further in this report.

\noindent{\bf Load Global (and other AUIPC-based idioms)}

The AUIPC instruction adds an immediate to the current PC address. 
Although similar to the use of the LUI instruction, AUIPC allows
for accessing data at arbitrary locations.

{\footnotesize
\begin{verbatim}
    // ld rd, symbol[31:0]
    auipc rd, symbol[31:12]
    ld    rd, symbol[11:0](rd)
\end{verbatim}
}
 
AUIPC is also used for jumping to routines more than 1 MB in distance
(AUIPC+JALR).  However, the AUIPC instruction is executed incredibly
rarely in SPECInt2006 given our compiler options in Table
\ref{table:cflags}, and so AUIPC idioms are not explored in this
report.  They will occur more frequently in dynamically linked code.

%
%
%
%

We note also that the RISC-V manual for the ``M'' multiply-divide
extension already indicates several idioms for multiply/divide
instruction pairings to enable microarchitectural fusing for wide
multiplies, to return both high and low words of a product in one
multiply, and for division, to return both quotient and remainder from
one division operation.

\subsection{Results of Macro-op Fusion}\label{sec:macroop-results}

Using the histogram counts and disassembly data from RV64GC
executions, we computed the number of macro-op fusion opportunities
available to RISC-V processors. This was a two-step process. The first
step was to automatically parse the instruction loops for fusion
pairs.  However, as the RISC-V {\tt gcc} compiler is not aware of
macro-op fusion, this automated process only finds macro-op fusion
pairs that exist serendipitously. The second step was to manually
analyze the 80\% most-executed loops of all 35 workloads for any
remaining macro-op fusion opportunities. The typical scenario involved
the compiler splitting apart potential fusion pairs with an unrelated
instruction or allocating a destination register that failed to
clobber the side-effect of the first instruction in the idiom pair.
This latter scenario required verifying that a clobber could have been
safely performed:

\begin{center}
\begin{minipage}{\textwidth}
\begin{lstlisting}[caption=A potential macro-op fusion opportunity from {\tt 403.gcc} ruined by oblivious register allocation. As the {\tt ld} is the last reader of {\tt a4} it can safely clobber it., label=code:failed-fusion]
add a4, a4, a5
ld  a3, 0(a4)
li  a4, 1
\end{lstlisting}
\end{minipage}
\end{center}

As 57\% of fusion pairs were found via the manual process, compiler
optimizations will be required to take full advantage of macro-op
fusion in RISC-V.


Figure \ref{fig:instcounts} shows the results of RV64GC macro-op
fusion relative to the other ISA instruction counts. Macro-op fusion
enables a 5.4\% reduction in {\em effective} instructions, allowing
RV64GC to execute 4.2\% fewer operations relative to \amd's {\em
  micro-op} count.

\setlength{\tabcolsep}{0.1em} 
\begin{table}[htp]
\caption{RISC-V RV64 Macro-op Fusion Opportunities.}
\begin{center}
\begin{tabular}{|c|cccc|}
\cline{3-5}
\multicolumn{2}{c}{} & \multicolumn{3}{|c|}{ \% reduction in effective instruction count}\\
\hline
& \multicolumn{1}{|c|}{macro-op to} &\multicolumn{1}{|c|}{\ \ \ indexed \ \ \ } & \multicolumn{1}{|c|}{\ \ clear upper \ \ } & \multicolumn{1}{|c|}{load effective} \\
benchmark       & \multicolumn{1}{|c|}{instruction} & \multicolumn{1}{|c|}{load}             & \multicolumn{1}{|c|}{word}    & \multicolumn{1}{|c|}{address} \\
                &  \multicolumn{1}{|c|}{ ratio }   &  \multicolumn{1}{|c|}{(add, ld)}   & \multicolumn{1}{|c|}{(slli, srli)} &  \multicolumn{1}{|c|}{(slli, add)} \\
\hline                                     
400.perlbench   & 0.97 &     1.27  & 0.12 & 1.59 \\
401.bzip2       & 0.81 &     8.55  & 5.58 & 4.67 \\
403.gcc         & 0.99 &     0.64  & 0.31 & 0.49 \\
429.mcf         & 0.99 &     0.38  & 0.00 & 0.31 \\
445.gobmk       & 0.94 &     3.62  & 0.14 & 2.61 \\
456.hmmer       & 1.00 &     0.01  & 0.01 & 0.02 \\
458.sjeng       & 0.90 &     5.01  & 0.01 & 4.88 \\
462.libquantum  & 1.00 &     0.00  & 0.00 & 0.01 \\
464.h264ref     & 0.89 &     5.39  & 0.02 & 5.70 \\
471.omnetpp     & 0.98 &     0.92  & 0.09 & 0.54 \\
473.astar       & 0.91 &     3.43  & 0.00 & 6.05 \\
483.xalancbmk   & 1.00 &     0.09  & 0.06 & 0.05 \\
\hline
arithmetic mean & 0.95 &     2.44  & 0.53 & 2.24 \\
\hline
\end{tabular}
\end{center}
\label{table:fusion}
\end{table}%

Table \ref{table:fusion} shows the breakdown of the different
SPECInt2006 workloads and the profitability of different
idioms. Although macro-op fusion provides an average of 5.4\%
reduction in effective instructions (in other words, 10.8\% of
instructions are part of a fusion pair), the variance between
benchmarks is significant: half of the benchmarks exhibit less than
2\% reduction while three experience a roughly 10\% reduction and {\tt
  401.bzip2} experiences a nearly 20\% reduction.

\subsection{A Design Proposal: Adding Macro-op Fusion to the Berkeley Rocket in-order core}

Macro-op fusion is not only a technique for high-performance
super-scalar cores.  Even single-issue cores with no compressed ISA
support like the RV64G 5-stage Rocket
processor~\cite{Asanovic:EECS-2016-17} can benefit.  To support
macro-op fusion, Rocket can be modified to fetch and decode up to two
4-byte instructions every cycle.\footnote{``Overfetching'' is actually quite advantageous as the \
access to the instruction cache is energy-expensive.  Indeed, for this \
exact reason, Rocket already overfetches up to 16 bytes at a time from \
the instruction cache and stores them in a buffer.}  If fusion is possible, the two
instructions are passed down the pipeline as a single {\em macro-op}
and the PC is incremented by 8 to fetch the next two instructions.  In
this manner, Rocket could reduce the latency of some idioms and
effectively execute fewer instructions by fusing them in the {\em
  decode} stage.
  
  Handling exceptions will require some care.
If the second instruction in a fusion pair causes an exception, the trap must be taken with the result of the first instruction visible in the architectural register file. This may be fairly straight forward for many micro-architectures such as Rocket - the value to be written back to the destination register can be changed to the intermediate value and the {\em Exception Program Counter} can be pointed to the second instruction. However, some implementations may find it easier to re-execute the pair in a ``don't fuse'' mode to achieve the correct behavior.

\subsection{Additional RISC-V Macro-op Fusion Pairs}

Although we only explore three fusion pairs in this report (as they should be relatively trivial -- and profitable -- for virtually all RISC-V pipelines), there are a number of other macro-op fusion pairs whose profitability will depend on the specific micro-architecture and benchmarks. A few are briefly discussed in this section.

\noindent{\bf Wide Multiply/Divide \& Remainder}

Given two factors of size $xlen$, a multiply operation generates a product of size $2*xlen$. 
In RISC-V, two separate instructions are required to get the full $2*xlen$ of the product - {\tt MULH} and {\tt MUL} (to get the high-order $xlen$ bits and the low-order $xlen$ bits separately). 

{\footnotesize
\begin{verbatim}
    MULH[[S]U] rdh, rs1, rs2
    MUL        rdl, rs1, rs2 
\end{verbatim}
}

In fact, the RISC-V user-level manually explicitly recommends this sequence as a fusion pair \cite{Waterman:EECS-2014-54}.
%
Likewise, the RISC-V user-level manual also recommends fusing divide/remainder instructions:

{\footnotesize
\begin{verbatim}
    DIV[U] rdq, rs1, rs2 
    REM[U] rdr, rs1, rs2 
\end{verbatim}
}

\noindent{\bf Load-pair/Store-pair}


ARMv8 uses load-pair/store-pair to read from (or write to) up to 128 contiguous bits in memory into (or from) two separate registers in a single instruction.
This can be re-created in RISC-V by fusing back-to-back loads (or stores) that read (or write) to contiguous addresses in memory.  Note that this can still be fused in decode as the address of the load does not need to be known, only that the pair of loads read the same base register and their immediates differ only by the size of the memory operation:

{\footnotesize
\begin{verbatim}
    // ldpair rd1,rd2, [imm(rs1)]
    ld    rd1, imm(rs1)
    ld    rd2, imm+8(rs1)
\end{verbatim}
}
 
As load-double is available in RVC, the common case of moving 128-bits can be performed by a single fused 4-byte sequence.  To make the pairing even easier  to detect, RVC also contains loads and stores that implicitly use the stack pointer as the common base address. In fact, register save/restore sequences are the dominant case of the load/store multiple idiom.

As discussed in Section \ref{sec:armv8}, load-pair instructions are not cheap - they require two write-ports on the register file.  Likewise, store-pair instructions require three read-ports.  In addition to the register file costs, processors with complex load/store units may suffer from additional complexity.
 
%

\noindent{\bf Post-indexed Memory Operations}


Post-indexed memory operations allow for a single instruction to perform a load (or store) from a memory address and then to increment the register holding the base memory address.

{\footnotesize
\begin{verbatim}
    // ldia rd, imm(rs1)
    ld    rd, imm(rs1)
    add   rs1, rs1, 8
\end{verbatim}
}

A couple of things are worth noting. First, both instructions are compressible,
allowing this fusion pair to typically fit into a single 4-byte sequence.
Second, two write-ports are required for post-indexed loads, making this fusion not profitable for
all micro-architectures.\footnote{Post indexed stores can use the existing write port, but if the increment is different from the offset, two adders are required. While the hardware cost may largely be regarded as trivial, a number of ISAs only support the {\tt ld~rd,~imm(rs1)++} form of address incrementing, including ARMv8.}

\section{ARMv8 Micro-op Discussion}\label{sec:armv8}

\setlength{\tabcolsep}{1.0em} 
\begin{table*}[htp]
\caption{ARMv8 memory instruction counts. Data is shown for normal loads (ld), loads with increment addressing (ldia), load-pairs (ldp), and load-pairs with increment addressing (ldpia). Data is also shown for the corresponding stores. Many of these instructions are likely candidates to be broken up into micro-op sequences when executed on a processor pipeline. For example, ldia and ldp require two write ports and the ldpia instruction requires three register write ports.}
\begin{center}
\begin{tabular}{|c|cccc|cccc|}
\hline
\multicolumn{1}{|c}{benchmark}  & \multicolumn{8}{|c|}{ \% of total ARMv8 instruction count}\\
\cline{2-9}
& \multicolumn{1}{|c|}{ld} & \multicolumn{1}{|c|}{ldia} & \multicolumn{1}{|c|}{ldp} & \multicolumn{1}{|c|}{ldpia} & \multicolumn{1}{|c|}{st} & \multicolumn{1}{|c|}{stia} & \multicolumn{1}{|c|}{stp} & \multicolumn{1}{|c|}{stpia} \\
\hline
400.perlbench   & 18.18 & 0.06  & 3.87  & 1.02  & 6.14  & 1.02  & 3.81  & 1.02  \\
401.bzip2   & 22.85 & 1.71  & 0.53  & 0.02  & 8.28  & 0.02  & 0.24  & 0.02  \\
403.gcc         & 16.80 & 0.11  & 2.89  & 1.04  & 3.32  & 1.04  & 3.03  & 1.04  \\
429.mcf         & 26.61 & 0.01  & 3.21  & 0.07  & 3.76  & 0.07  & 3.22  & 0.07  \\
445.gobmk   & 15.77 & 1.01  & 2.04  & 0.77  & 6.14  & 0.74  & 2.19  & 0.74  \\
456.hmmer   & 24.20 & 0.09  & 0.06  & 0.02  & 13.75 & 0.02  & 0.01  & 0.02  \\
458.sjeng   & 17.37 & 0.00  & 1.30  & 0.26  & 4.38  & 0.26  & 1.46  & 0.26  \\
462.libquantum  & 14.00 & 0.00  & 0.15  & 0.06  & 1.85  & 0.06  & 0.31  & 0.06  \\
464.h264ref & 28.36 & 0.01  & 6.61  & 1.85  & 3.18  & 1.82  & 5.91  & 1.82  \\
471.omnetpp & 19.16 & 0.45  & 2.56  & 1.55  & 8.43  & 1.54  & 3.11  & 1.54  \\
473.astar   & 24.08 & 0.01  & 0.84  & 0.15  & 3.73  & 0.15  & 0.83  & 0.15  \\
483.xalancbmk   & 20.94 & 4.84  & 1.82  & 0.68  & 1.74  & 0.67  & 1.51  & 0.67  \\
\hline
arithmetic mean & 20.69 & 0.69  & 2.16  & 0.62  & 5.39  & 0.62  & 2.14  & 0.62  \\
\hline
\end{tabular}
\end{center}
\label{table:arm-qemu-data}
\end{table*}

\begin{figure}[htb]
        \centering
{\includegraphics[scale=1.0]{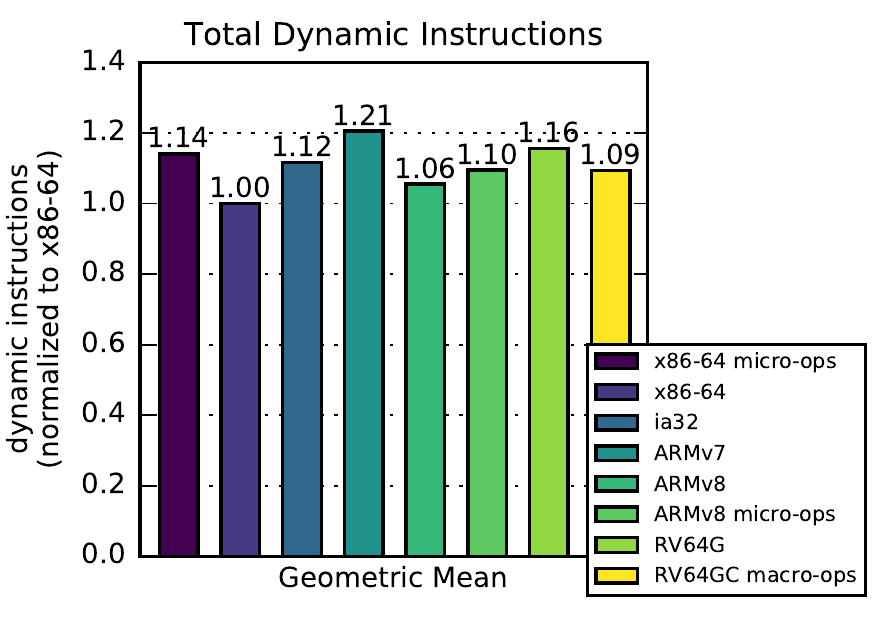}}
        \caption{
            The geometric mean of the instruction counts of the twelve SPECInt
            benchmarks is shown for each of the ISAs, normalized to \amd.  The \amd\ micro-op count is reported from the micro-architectural counters on an Intel Ivy Bridge processor.  The RV64GC macro-op count was collected as described in Section \ref{sec:macroop-results}. The ARMv8 micro-op count was synthetically created by breaking up {\em load-increment-address}, {\em load-pair}, and {\em load-pair-increment-address} into multiple micro-ops.
        }
        \label{fig:instcounts_geomean}
\end{figure}

As shown in Figure \ref{fig:instcounts}, the ARMv8 {\tt gcc} compiler emits 9\% fewer instructions than RV64G.  However, the ARMv8 instruction count is not necessarily an accurate measure of the amount of ``work'' an ARMv8-compatible processor must perform.
ARMv8 is implemented on a wide range of micro-architectures; each design may make different decisions on 
how to map the ARMv8 ISA to its particular pipeline. 
The Cortex-A53 processor used in this report does not provide a {\em retired micro-op} counter, so we must make an educated guess as to how a reasonable ARMv8 processor would break each ISA instruction into micro-ops.

Figure \ref{fig:instcounts_geomean} shows a summary of the total dynamic instruction count of the different ISAs, as well as the effective operation counts of \amd, RV64GC, and our best guess towards the ARMv8 micro-op count.
To generate our synthetic ARMv8 micro-op count, we assumed that any instruction that writes multiple registers would be broken down into additional micro-ops (one micro-op per register write-back destination).  


Table \ref{table:arm-qemu-data} provides the details behind our synthetic ARMv8 micro-op count. We first chose a set of ARMv8 instructions that are likely candidates for being broken up into multiple micro-ops.
In particular, ARMv8 supports memory operations with increment addressing modes and load-pair/store-pair instructions.
Two write-ports are required for the load-pair instruction (ldp) and for loads with
increment addressing (ldia), while three write-ports are required for
load-pair with increment addressing (ldpia). 
We then modified the QEMU ARMv8 ISA simulator to count these instructions that
are likely candidates for generating multiple micro-ops.

Although we show the breakdown of all load and store instructions in Table \ref{table:arm-qemu-data}, we assume for Figure \ref{fig:instcounts} that only ldia, ldp, and ldpia increase the micro-op count for our hypothetical ARMv8 processor.  Cracking these instructions into multiple micro-ops leads to an average increase of 4.09\%  in the operation count for ARMv8.
As a comparison, the Cortex-A72 out-of-order processor is reported to emit ``less than 1.1
micro-ops'' per instruction and breaks down ``move-and-branch'' and
``load/store-multiple'' into multiple micro-ops~\cite{cortexa72}.

We note that it is possible to ``brute-force'' these ARMv8 instructions and
handle them as a single operation within the processor backend. 
Many ARMv8 integer instructions require
three read ports, so it is likely that most (if not all) ARMv8 cores will pay
the area overhead of a third read port for the complex store instructions.  
Likewise, they can pay the cost to add a second (or even third) write port 
to natively support the load-pair and increment addressing modes.  
Of course, there is nothing that prevents a RISC-V core from taking on this
complexity, adding the additional register ports, and using macro-op fusion to
emulate the same complex idioms that ARMv8 has chosen to declare at the ISA level.  

\section{Recommendations}

A number of lessons can be learned from analyzing RISC-V's performance on SPECInt.

\subsection{Programmers}

Although it is not legal to modify SPEC for benchmarking, an analysis of its
hot loops highlight a few coding idioms that can hurt performance on RISC-V (and
often other) platforms.

\begin{itemize}
\item Avoid unsigned 32-bit integers for array indices. The {\tt
  size\_t} type should be used for array indexing and loop counting.
\item Avoid multi-dimensional arrays if the sizes are known and
  fixed. Each additional dimension in the array is an extra level of
  indirection in C, which is another load from memory.
\item C standard aliasing rules can prevent the compiler from making
  optimizations that are otherwise ``obvious'' to the programmer.  For
  example, you may need to manually `lift' code out of a loop that
  returns the same value every iteration.
\item Use the {\tt -fno-tree-loop-if-convert} flag to {\tt gcc} to disable a
problematic optimization pass that generates poor code.
\item Profile your code. An extra, unnecessary instruction in a hot
  loop can have dramatic effects on performance.
\end{itemize}

\subsection{Compiler Writers}

Table \ref{table:fusion} shows that a significant amount of potential
macro-op fusion opportunities exist, but relying on serendipity leaves
over half of the performance on the table. Any pursuit of macro-op
fusion in a RISC-V processor will require modifying the compiler to
increase the amount of fuse-able pairs in compiler-generated code.

The good news is that the {\tt gcc} compiler already supports an
instruction scheduling hook for macro-op fusion.\footnote{The {\tt
    gcc} back-end uses {\tt TARGET\_SCHED\_MACRO\_FUSION\_PAIR\_P
    (rtx\_insn *prev, rtx\_insn *curr)} to query if two instructions
  are a fuse-able pair.}  However, macro-op fusion also requires a
different register allocation scheme that aggressively overwrites
registers once they are no longer live, as shown in Code
\ref{code:failed-fusion}.

Finally, there will always be more opportunities to improve the code
scheduling. In at least one critical benchmark ({\tt 462.libquantum}),
store data generation was lifted outside of an inner branch and
executed every iteration, despite the actual store being gated off by
the condition and rarely executed. That one change would reduce the
RISC-V instruction count by 10\%!


\subsection{Micro-architects}

Macro-op fusion is a potentially quite profitable technique to
decrease the effective instruction count of programs and improve
performance.  What constitutes the set of profitable idioms will depend
significantly on the benchmark and the target processor pipeline.  For
example, in-order processors may be far more amenable to fusions that
utilize multiple write-back destinations (e.g., post-indexed memory
operations).  
When macro-op fusion is implemented along with other micro-architectural techniques such as micro-op caches and loop buffers, 
the ISA instruction count of a program can be much greater than the effective instruction count.


\section{Future Work}

This work is just the first step in continuing to evaluate and assess the quality of the RISC-V code generation.  Future work should look at new benchmarks and new languages. In particular,  {\em just-in-time (JIT)} and {\em managed}
languages may exhibit different behaviors, and thus favor different
idioms, than the C and C++ code used by the SPECInt benchmark
suite~\cite{blackburn2008wake}.
Even analyzing SPECfp, which includes benchmarks written in Fortran, 
would explore a new dimension of RISC-V.
Unfortunately, much of the future work is predicated
on porting and tuning new run-times to RISC-V.
\section{Conclusion}\label{sec:conclusion}

Our analysis using the SPEC CINT2006 benchmark suite shows that the 
RISC-V ISA can be both denser and higher performance than the 
popular, existing commercial CISC ISAs.
In particular, the RV64G ISA on
average executes 16\% more instructions per program than \amd\ and
fetches 23\% more instruction bytes.  When coupled with the RISC-V
Compressed ISA extension, the dynamic instruction bytes per program
drops significantly, helping RV64GC fetch 8\% fewer instruction bytes
per program relative to \amd.  Finally, an RV64 processor that
supports macro-op fusion, coupled with a fusion-aware
compiler, could see a 5.4\% reduction in its ``effective'' instruction
count, helping it to execute 4.2\% fewer effective instructions
relative to \amd's micro-op count.

There are many reasons to keep an instruction set elegant and simple, especially for a free and open ISA. Macro-op fusion allows per implementation tuning of the effective ISA without burdening subsequent generations of processors with optimizations that may not make sense for future programs, languages, and compilers. It also allows asymmetric optimizations that are anathema to compiler writers and architectural critics. 

Macro-op fusion has been previously used by commercial ISAs like ARM 
and x86 to accelerate idioms created by legacy ISA decisions 
like the two-instruction {\em compare-and-branch} sequence.
For RISC-V, we have the power to change the ISA, but it is actually better not to! 
Instead, we can leverage macro-op fusion in a new way -- to specialize processors to their designed tasks, while leaving the ISA -- which must try to be all things to all people -- unchanged.  



 \section*{Acknowledgments}
 
 The authors would like to thank Scott Beamer, Brian Case, David Ditzel, and Eric Love for their valuable feedback.

Research partially funded by DARPA Award Number HR0011-12-2-0016, the
Center for Future Architecture Research, a member of STARnet, a
Semiconductor Research Corporation program sponsored by MARCO and
DARPA, and ASPIRE Lab industrial sponsors and affiliates Intel,
Google, HPE, Huawei, LGE, Nokia, NVIDIA, Oracle, and Samsung.  Any
opinions, findings, conclusions, or recommendations in this paper are
solely those of the authors and does not necessarily reflect the
position or the policy of the sponsors.

\

\bstctlcite{bstctl:etal, bstctl:nodash, bstctl:simpurl}
\bibliographystyle{IEEEtranS}
\bibliography{IEEEabrv,references}

\vfill

\pagebreak

\appendix

This section covers in more detail the behavior of some of the most commonly executed loops for SPECInt 2006.
More information about individual benchmarks can be found at 

\noindent \url{http://www.spec.org/cpu2006/docs/}.

This section also includes the raw dynamic instruction counts used in this study, shown in Table \ref{table:rawinstcounts}.
 
\setlength{\tabcolsep}{0.75em} 
\begin{table*}[htp]
\caption{Total dynamic instructions (in billions) when compiled using {\tt gcc 5.3 -O3 -static}. The \amd\ {\em retired micro-op count} is also shown as measured using an Intel Xeon (Ivy Bridge).}
\begin{center}
\begin{tabular}{|c|rrrrrr|}
\hline
    benchmark      & \multicolumn{1}{c}{\amd\ uops} & \multicolumn{1}{c}{\amd} & \multicolumn{1}{c}{IA-32} &  \multicolumn{1}{c}{ARMv7} & \multicolumn{1}{c}{ARMv8} & \multicolumn{1}{c|}{RV64G} \\
\hline
    400.perlbench  &  2,367.2 &  2,091.4 &  2,170.9 &  2,436.2 &  2,229.9 &  2,446.9 \\
    401.bzip       &  2,523.7 &  2,260.2 &  2,372.9 &  2,340.8 &  2,339.1 &  3,006.7 \\
    403.gcc        &  1,143.1 &    963.6 &    997.1 &  1,246.6 &    939.3 &  1,313.3 \\
    429.mcf        &    305.2 &    294.2 &    315.7 &    350.7 &    300.0 &    276.8 \\
    445.gobmk      &  1,825.4 &  1,645.8 &  1,651.6 &  1,961.3 &  1,812.6 &  1,947.0 \\
    456.hmmer      &  3,700.7 &  2,525.7 &  2,996.2 &  3,665.2 &  3,057.4 &  2,929.0 \\
    458.sjeng      &  2,376.1 &  2,223.2 &  2,359.4 &  2,714.0 &  2,494.1 &  2,872.3 \\
    462.libquantum &  1,454.8 &  1,649.1 &  2,675.0 &  2,274.5 &  1,562.7 &  1,365.6 \\
    464.h264ref    &  4,348.9 &  2,952.8 &  3,054.0 &  3,440.5 &  3,357.9 &  4,841.4 \\
    471.omnetpp    &    687.7 &    553.1 &    661.7 &    599.2 &    544.4 &    578.6 \\
    473.astar      &    989.4 &    949.0 &  1,055.6 &  1,114.2 &    997.6 &    936.9 \\
    483.xalancbmk  &    926.1 &    864.0 &    951.7 &  1,019.4 &    906.3 &    990.3 \\
\hline
\end{tabular}
\end{center}
\label{table:rawinstcounts}
\end{table*}

\subsection{400.perlbench}

{\tt 400.perlbench} benchmarks the interpreted Perl language with some of the
more OS-centric elements removed and file I/O reduced.  

Although {\tt libc\_malloc} and {\tt \_int\_free} routines make an appearance for a few percent of the instruction count, the only thing of any serious note in {\tt 400.perlbench} is a significant amount of the instruction count spent on stack pushing and popping. This works against RV64G as its larger register pool requires it to spend more time saving and restoring more registers. Although counter-intuitive, this can be an issue if functions exhibit early function returns and end up not needing to use all of the allocated registers.

\subsection{401.bzip2}

{\tt 401.bzip2} benchmarks the bzip2 compression tool, modified to perform the
compression and decompression entirely in memory.

\begin{center}
\begin{minipage}[tbp]{1.0\textwidth}
\begin{lstlisting}[caption=The mainSort routine in 401.bzip. Line 7 accounts for $>$3\% of the RV64G instruction count., label=code_bzip]
// UInt32* ptr
// UChar*  block
// UInt16* quadrant
// UInt32* ftab
// Int32   unHi

n = ((Int32)block[ptr[unHi]+d]) - med;


// RV64G assembly for line 7
35a58:   lw      a4, 0(t4)
35a5c:   addw    a5, s3, a4
35a60:   slli    a5, a5, 0x20
35a64:   srli    a5, a5, 0x20
35a68:   add     a5, s0, a5
35a6c:   lbu     a5, 0(a5)
35a70:   subw    a5, a5, t3
35a74:   bnez    a5, 35b00


// x86-64 assembly for line 7
4039d0:  mov    (%r10), %edx
4039d3:  lea    (%r15,%rdx,1), %eax
4039d7:  movzbl (%r14,%rax,1), %eax
4039dc:  sub    %r9d, %eax
4039df:  cmp    $0x0, %eax
4039e2:  jne    403a8a
\end{lstlisting}
\end{minipage}
\end{center}

Aside from the {\tt 403.gcc} (memset) and {\tt 464.h264ref} (memcpy)
benchmarks, {\tt 401.bzip2} is RV64G's worst performing benchmark. {\tt
401.bzip2} spends a significant amount of instructions manipulating arrays
using {\em unsigned} 32-bit integers.  Code \ref{code_bzip} shows that the
index into the {\tt block} array is an {\em unsigned} 32-bit integer. As RISC-V
does not have unsigned arithmetic, and the RV64 ABI specifies that the 64-bit
registers stores {\em signed} values, extra instructions are required to clear
the upper 32-bits of the index variable before the load access can be
performed. This behavior is consistent with the MIPS and Alpha ISAs. On the
other hand, ARMv8 and x86-64 provide addressing modes that only read (or write to)
parts of the full 64-bit register.

As the majority of \bzip\ is composed of array accesses, it is little surprise
that RISC-V's lack of indexed loads, load effective address, and low word
accesses translates to 33\% more RV64G instructions relative to \amd. However,
when using macro-op fusion, nearly 40\% of instructions can be combined to
reduce the effective instruction count by 20\%, which puts RV64G as using 3\% fewer operations than
the \amd\ micro-op count for \bzip.

\subsection{403.gcc}
 
\begin{center}
\begin{minipage}{1.0\textwidth}
\begin{lstlisting}[caption=The memset routine in 403.gcc., label=code_gcc_memset]
// RV64G, 4 instructions to move 16 bytes
4a3814:    sd      a1, 0(a4)
4a3818:    sd      a1, 8(a4)
4a381c:    addi    a4, a4, 16
4a3820:    bltu    a4, a3, 4a3814


// x86-64, 7 instructions to move 64 bytes
6f24c0:    movdqa  %xmm8, (%rcx)
6f24c5:    movdqa  %xmm8, 0x10(%rcx)
6f24cb:    movdqa  %xmm8, 0x20(%rcx)
6f24d1:    movdqa  %xmm8, 0x30(%rcx)
6f24d7:    add     $0x40, %rcx
6f24db:    cmp     %rcx, %rdx
6f24de:    jne     6f24c0


// armv8, 6 instructions to move 64 bytes
6f0928:    stp     x7, x7, [x8,#16]
6f092c:    stp     x7, x7, [x8,#32]
6f0930:    stp     x7, x7, [x8,#48]
6f0934:    stp     x7, x7, [x8,#64]!
6f0938:    subs    x2, x2, #0x40
6f093c:    b.ge    6f0928
\end{lstlisting}
\end{minipage}
\end{center}
  
{\tt 403.gcc} benchmarks the {\tt gcc 3.2} compiler generating code for
the \amd\ AMD Opteron processor.  Despite
being a ``SPECInt'' benchmark, {\tt 403.gcc} executes an optimization pass that
performs constant propagation of floating-point constants, which requires IEEE
floating-point support and can lead to significant execution time spent in
soft-float routines if hardfloat support is not available.
 
30\% of RISC-V's instruction count is devoted to the {\tt memset} routine. The
critical loop for {\tt memset} is shown in Code \ref{code_gcc_memset}. ARMv8
and \amd\ use a single instruction to move 128 bits. Their critical loop is
also unrolled to better amortize the loop bookkeeping instructions. ARMv8 is an
instruction shorter than \amd\ as it rolls the address update into one of its
``store-pair'' post-indexing instructions.

\subsection{429.mcf}

{\tt 429.mcf} executes a routine for scheduling bus routes (``network flows'').
The core routine is an implementation of {\tt simplex}, an optimization
algorithm using linear programming. The performance of {\tt 429.mcf} is
typically memory-bound.

RV64G emits the fewest instructions of all of the tested ISAs.
For RV64G, the top 31\% of {\tt 429.mcf} is contained in  just 14 instructions - and five of
those instructions are branches.
The other ISAs typically require two instructions to describe a conditional branch, explaining their higher instruction counts.

\subsection{445.gobmk}

{\tt 445.gobmk} simulates an AI analyzing a Go board and suggesting
moves. Written in C, it relies significantly on structs (and macros)
to provide a quasi-object-oriented programming style. This translates
to a significant number of indexed loads which penalizes RV64G's
instruction count relative to other ISAs.

A {\tt memset} routine makes up around 1\% of the benchmark, in
which \amd\ leverages a {\tt movdqa} instruction to write 128 bits at
a time.

An example of sub-optimal RV64G code generation is shown in
Code \ref{code_gobmk}.  Although only one conditional {\em if}
statement is described in the C code to guard assignments to two
variables ({\tt smallest\_dist} and {\tt best\_index}), the compiler
emits two separate branches (one for each variable). Compounding on
this error, the two variables are shuttled between registers {\tt t4}
and {\tt t6} and {\tt a0} and {\tt a3} three separate times each.

\begin{center}
\begin{minipage}{1.0\textwidth}
\begin{lstlisting}[caption={Sub-optimal RV64G code generation in 445.gobmk, accounting for 3.5\%.}, label=code_gobmk]
// smallest_dist = 10000

/* Find the smallest distance among the queued points. */
for (k = conn->queue_start; k < conn->queue_end; k++) {
  if (conn->distances[conn->queue[k]] < smallest_dist) {
    smallest_dist = conn->distances[conn->queue[k]];
    best_index = k;
  }
}

// RV64G assembly
<compute_connection_distances>
...
550ef8:    addi    a5, a4, 2000
550efc:    slli    a5, a5, 0x2
550efe:    add     a5, a5, s8
550f00:    lw      a5, 0(a5)      // conn->queue[k]  
550f02:    mv      t6, a4         // ??
550f04:    mv      a0, a3         // ??
550f06:    slli    a5, a5, 0x2
550f08:    add     a5, a5, s8
550f0a:    lw      a5, 0(a5)      // conn->distances[conn->queue[k]]
550f0c:    addiw   a4, a4, 1
550f0e:    ble     a3, a5, 550f14 // first branch, for smallest_dist
550f12:    mv      a0, a5

550f14:    blt     a5, a3, 550f1a // ?!?!
550f18:    mv      t6, t4

550f1a:    mv      t4, t6         // ?? 
550f1c:    mv      a3, a0         // ??
550f1e:    bne     a4, a2, 550ef8
\end{lstlisting}
\end{minipage}
\end{center}

The bad code generation can be rectified by turning off the
{\tt tree-loop-if-convert} optimization pass.  The compiler attempts to
use conditional moves to remove branches in the inner-most branch to
facilitate vectorization, but as RISC-V lacks conditional move
instructions, this optimization pass instead interferes with the other
passes and instead, as a final step, emits a poor software imitation
of a conditional move for each variable assignment.  By using the
{\tt -fno-tree-loop-if-convert} flag to gcc, the total instruction count
of {\tt 445.gobmk} is reduced by 1.5\%.



\subsection{456.hmmer}

{\tt 456.hmmer} benchmarks a hidden Markov model searching for patterns in DNA
sequences. Nearly 100\% of the benchmark is contained within just 70
instructions, all within the optimized {\tt P7Viterbi} function.

RV64G outperforms all other ISAs with the exception of \amd. The {\tt
P7Viterbi} function contains a significant number of short branches
around a store with the branch comparison typically between array
elements. For \amd, the load from memory and the comparison can be
rolled into a single instruction.

\begin{center}
\begin{minipage}{1.0\textwidth}
\begin{lstlisting}[caption=An example of a typical idiom from P7Viterbi function in 456.hmmer., label=code_hmmer]
if ((sc = ip[k-1]  + tpim[k-1]) > mc[k])  mc[k] = sc;
\end{lstlisting}
\end{minipage}
\end{center}

Due to \amd's CISC memory addressing modes, even `simple' instructions like {\tt add} can become quite expressive:

{\footnotesize
\begin{verbatim}
    add 0x4(%rbx,%rdx,4),%eax
\end{verbatim}}

This is a common instruction in \hmmer\ which describes a shift, a
register-register add, a register-immediate add, a load from memory,
and a final addition between the load data and the {\tt eax} register.
However, despite \amd's lower instruction count (due largely to the
memory addressing modes), the \amd\ retired micro-op count is 26\%
more than the RV64G instruction count.

As an interesting final note, the end of the {\tt P7Viterbi} function
contains an expensive floating-point divide by 1000.0, to scale the
integer scores to floating-point scores at the end of an integer
arithmetic routine.  Although rarely executed, this can be punishing
for micro-architectures that do not support floating-point divide in
hardware.\footnote{This floating-point divide can be quite a
surprising find in the SPEC {\bf Integer} benchmark suite. Although it
is rarely executed, the cost to emulate it in software (and its
neighbors {\tt fcvt} and {\tt flw}) can become noticeable.}

\subsection{458.sjeng}

{\tt 458.sjeng} benchmarks an AI playing Chess using {\em alpha-beta} tree
searches, game-board evaluations, and pruning.

The following section of code demonstrates a lost potential fusion
opportunity which shows the importance of a more intelligent compiler
register allocation scheme.  By using register {\tt t1} in line 2, the
add/lw pair cannot be fused as the side-effect to {\tt t1} must remain
visible. A better implementation would use {\tt a1} in place of {\tt
t1}, which would allow the add/lw pair to be fused as the side-effect
from the add will now be clobbered. Note that {\tt t1} is clobbered in
line 5, so the proposed transformation is safe.

\begin{center}
\begin{minipage}{1.0\textwidth}
\begin{lstlisting}[caption={A missed fusion opportunity in 458.sjeng due to the register allocation.}, label=code_label]
// currently emitted code:
   add     t1, s2, s3
   lw      a1, 0(t1)
   sw      a0, 80(sp)
   li      t1, 12
   addiw   t3, a2, 1
   bltu    t1, t6, 2ba488

// proposed fusible version:
   add     a1,s2,s3
   lw      a1,0(a1)
   sw      a0,80(sp)
   li      t1,12
   addiw   t3,a2,1
   bltu    t1,t6, 2ba488
\end{lstlisting}
\end{minipage}
\end{center}

\subsection{462.libquantum}

{\tt 462.libquantum}  simulates a quantum computer executing Shor's algorithm.
On RV64G, 80\% of the dynamic instructions is spent on 6 instructions, and 90\%
is spent on 18 instructions. On \amd, 11 instructions account for 88\% of the
dynamic instructions. The hot loop is simulating a Toffoli gate.

Although RV64G emits fewer instructions for {\tt libquantum} relative to all
other ISAs, sub-optimal code is still being generated.  The store data
generation instruction ({\tt xor a4,a4,a2}) is executed every iteration,
regardless of the outcome of the inner-most branch.  Moving that
instruction inside the conditional with its store will save 10\% on
the dynamic instruction count!  It is possible the compiler is
attempting to generate a conditional-store idiom (a forward branch
around a single instruction). This is a potential macro-op
fusion opportunity for RISC-V pipelines that support conditional
moves, but is otherwise an extra, unnecessary instruction for all
other micro-architectures.


%

\begin{center}
\begin{minipage}{1.0\textwidth}
\begin{lstlisting}[caption=The hot loop for 462.libquantum., label=code_libquantum]
// int control1, cintrol2
for(i=0; i<reg->size; i++)
{
    /* Flip the target bit of a basis state if both control bits are set */
    if(reg->node[i].state & ((MAX_UNSIGNED) 1 << control1))
    {
        if(reg->node[i].state & ((MAX_UNSIGNED) 1 << control2))
        {
            reg->node[i].state ^= ((MAX_UNSIGNED) 1 << target);
        }
    }
}

<quantum_toffoli>:

// the conditional store resides in a rarely-true if condition

// RV64GC assembly
36ee6:    ld   a4, 0(a5)
36ee8:    and  a0, a4, a1
36eec:    xor  a4, a4, a2
36eee:    bne  a0, a1, 36ef4
36ef2:    sd   a4, 0(a5)

36ef4:    addi a5, a5, 16
36ef6:    bne  a3, a5, 36ee6


// ARMv7 assembly
1111c:    ldrd   r2, [ip, #8]
11120:    and    r5, r3, r1
11124:    and    r4, r2, r0
11128:    cmp    r5, r1
1112c:    eor    r2, r2, r8
11130:    cmpeq  r4, r0
11134:    eor    r3, r3, r9
11138:    strdeq r2, [ip, #8]
1113c:    add    ip, ip, #16
11140:    cmp    ip, lr
11144:    bne    1111c


// ARMv8 assembly
4029b0:   ldr    x0, [x3]
4029b4:   bics   xzr, x1, x0
4029b8:   eor    x0, x0, x2
4029bc:   b.ne   4029c4
4029c0:   str    x0, [x3]

4029c4:   add    x3, x3, #0x10
4029c8:   cmp    x4, x3
4029cc:   b.ne   4029b0


// x86-64 assembly
401eb0:   mov    (%rax), %rdx
401eb3:   mov    %rdx, %rcx
401eb6:   and    %rsi, %rcx
401eb9:   cmp    %rsi, %rcx
401ebc:   jne    401ec4
401ebe:   xor    %r8, %rdx
401ec1:   mov    %rdx, (%rax)

401ec4:   add    $0x10, %rax
401ec8:   cmp    %rax, %rdi
401ecb:   jne    401eb0
\end{lstlisting}
\end{minipage}
\end{center}
 
The ARMv7 performance deviates significantly on this benchmark. The hot-path is
6 instructions for RISC-V and 11 for ARMv7. The
assembly code is shown Code \ref{code_libquantum}. The first point of interest
is the compiler uses a conditional store instruction instead of a branch. While
this can be quite profitable in many cases (it can reduce pressure on the
branch predictor), this particular branch is heavily biased to be {\em not-taken}
causing an extra instruction to be executed every iteration.
It also appears the compiler failed to coalesce the two branches together causing an extra three instructions to be emitted.
Finally, all ARM branches are a two-instruction idiom requiring an extra {\em
compare} instruction to set-up the condition code for the {\em branch-on-condition-code} instruction.

This poor code generation from ARMv7 is rectified in ARMv8, which is essentially identical to the RV64G code (modulo the extra instruction for branching).

Finally, we would be remiss to not mention that this loop is readily amenable
to vectorization. Each loop iteration is independent and a single conditional
affects whether the element store occurs or not.  With proper coaxing from
the Intel {\tt icc} compiler, an Intel Xeon can demonstrate a stunning 10,000x
 performance improvement on {\tt libquantum} over the baseline SPEC
machine (the geometric mean across the other benchmarks is typically 35-70x for Intel Xeons).

\subsection{464.h264ref}

The {\tt 464.h264ref} benchmark is a reference implementation of the
h264 video compression standard. 25\% of the RV64G dynamic
instructions is devoted to a {\tt memcpy} routine. It features a
significant number of multi-dimensional arrays that forces extra loads
to find the address of the actual array element.

Within the {\tt memcpy} routine, the ARMv7 code exploits load-multiple/store-multiple instructions to
move eight 32-bit registers of data per memory instruction (32 bytes per
loop iteration). The {\tt ldm/stmia} instructions also auto-increment
the base address source operand. ARMv8 has no
load-multiple/store-multiple and instead relies on
load-pair/store-pair to move eight registers in a 10 instruction loop.
However, the registers are twice as wide (32 bits versus 64 bits),
allowing ARMv8 to make up some ground at having lost the
load/store-multiple instructions.\footnote{One potential advantage of
load/store pair instructions over the denser load/store-multiple is
that it is possible to implement load/store pair as a single micro-op
at the cost of more register file ports.}
 
RV64G lacks any complex memory instructions, and instead emits a simple
unrolled sequence of 21 instructions that moves 72 bytes.  Meanwhile, \amd\
uses a single {\tt rep movsq} (repeat move 64-bits) instruction to
execute 60\% fewer instructions relative RV64G.

\begin{center}
\begin{minipage}{1.0\textwidth}
\begin{lstlisting}[caption=The memcpy loop for 464.h264ref., label=code_h264]
// RV64G
15a468:    ld    t2, 0(a1)
15a46c:    ld    t0, 8(a1)
15a470:    ld    t6, 16(a1)
15a474:    ld    t5, 24(a1)
15a478:    ld    t4, 32(a1)
15a47c:    ld    t3, 40(a1)
15a480:    ld    t1, 48(a1)
15a484:    ld    a2, 56(a1)
15a486:    addi  a1, a1, 72
15a48a:    addi  a4, a4, 72
15a48e:    ld    a3, -8(a1)
15a492:    sd    t2, -72(a4)
15a496:    sd    t0, -64(a4)
15a49a:    sd    t6, -56(a4)
15a49e:    sd    t5, -48(a4)
15a4a2:    sd    t4, -40(a4)
15a4a6:    sd    t3, -32(a4)
15a4aa:    sd    t1, -24(a4)
15a4ae:    sd    a2, -16(a4)
15a4b2:    sd    a3, -8(a4)
15a4b6:    bltu  a4, a5, 15a468


// x86-64
4ec93b:    rep movsq %ds:(%rsi), %es:(%rdi)


// ARMv7
e1174:     pld   [r1, #124]  ; 0x7c
e1178:     ldm   r1!, {r3, r4, r5, r6, r7, r8, ip, lr}
e117c:     subs  r2, r2, #32
e1180:     stmia r0!, {r3, r4, r5, r6, r7, r8, ip, lr}
e1184:     bge   e1174
e1188:     cmn   r2, #96 ; 0x60
e118c:     bge   e1178


// ARMv8
4ca314:    stp   x7, x8, [x6,#16]
4ca318:    ldp   x7, x8, [x1,#16]
4ca31c:    stp   x9, x10, [x6,#32]
4ca320:    ldp   x9, x10, [x1,#32]
4ca324:    stp   x11, x12, [x6,#48]
4ca328:    ldp   x11, x12, [x1,#48]
4ca32c:    stp   x13, x14, [x6,#64]!
4ca330:    ldp   x13, x14, [x1,#64]!
4ca334:    subs  x2, x2, #0x40
4ca338:    b.ge  4ca314
\end{lstlisting}
\end{minipage}
\end{center}

\subsection{471.omnetpp}

{\tt 471.omnetpp} performs a discrete event simulation of an Ethernet
network.  It makes limited use of the {\tt strcmp} routine (less than
2\%).

Integer-only processors looking to benchmark their SPECInt performance
may be in for a surprise - the most executed loops of {\tt
471.omnetpp}, accounting for 10\% of the RV64G instruction count,
involves floating-point operations and floating-point comparisons!

RV64G emits 4.6\% more instructions than \amd- about 30 billion more
instructions.  Although {\tt 471.omnetpp} is fairly branch heavy, many
of the branch comparisons are performed between memory locations,
allowing \amd\ to combine the load and the branch comparison into a
single instruction. Thus, both RV64G and \amd\ require two
instructions to perform a memory load, compare the data to a value in
another register, and branch on the outcome.

\begin{center}
\begin{minipage}{1.0\textwidth}
\begin{lstlisting}[caption={The most executed segment in 471.omnetpp (3.5\% for RV64G). RV64G is spending extra instructions to compute the address of the FP value it will compare for a branch.}, label=code_omnetpp]
 <_ZN12cMessageHeap7shiftupEi+0x52>
// RV64G assembly
ce6ee:     slli   a2, a1, 0x3
ce6f2:     add    a3, a3, a2
ce6f4:     ld     a3, 0(a3)
ce6f6:     fld    fa5, 144(a3)
ce6f8:     flt.d  a4, fa5, fa4
ce6fc:     bnez   a4, ce720


// x86-64 assembly
464571:    movslq   %esi, %r10
464574:    mov      (%r8,%r10,8), %rdx
464578:    vmovsd   0x90(%rdx), %xmm1
464580:    vucomisd %xmm1, %xmm0
464584:    ja       4645a8
\end{lstlisting}
\end{minipage}
\end{center}

\begin{center}
\begin{minipage}{1.0\textwidth}
\begin{lstlisting}[caption={A section of the x86-64 \_\_strcmp\_sse3 routine, accounting for 1.36\% of the total instruction count.}, label=code_omnetpp_strcmp]
564b96:    movlpd   (%rdi), %xmm1
564b9a:    movlpd   (%rsi), %xmm2
564b9e:    movhpd   0x8(%rdi), %xmm1
564ba3:    movhpd   0x8(%rsi), %xmm2
564ba8:    pxor     %xmm0, %xmm0
564bac:    pcmpeqb  %xmm1, %xmm0
564bb0:    pcmpeqb  %xmm2, %xmm1
564bb4:    psubb    %xmm0, %xmm1
564bb8:    pmovmskb %xmm1, %edx
564bbc:    sub      $0xffff, %edx
564bc2:    jne      565db0
\end{lstlisting}
\end{minipage}
\end{center}

\begin{center}
\begin{minipage}{1.0\textwidth}
\begin{lstlisting}[caption={A section of the RV64GC and ARMv8 strcpy routine.
    ARMv8 is one less instruction thanks to some clever addressing (the load
    and store share the same base register) and the use of an indexed
    store. However, RISC-V, coupled with macro-op fusion, could use the same technique to improve its own performance.}, label=code_omnetpp_strcpy_armv8]
// RV64GC assembly
<strcpy>:
...
172b04:    lbu     a4,0(a1)
172b08:    addi    a5,a5,1
172b0a:    addi    a1,a1,1
172b0c:    sb      a4,-1(a5)
172b10:    bnez    a4,172b04
172b12:    ret

// ARMv8 assembly
 <strcpy>:
54e5c0:    sub    x3, x0, x1
54e5c4:    ldrb   w2, [x1]
54e5c8:    strb   w2, [x1,x3]
54e5cc:    add    x1, x1, #0x1
54e5d0:    cbnz   w2, 54e5c4
54e5d4:    ret

// Here's a better RV64GC strcpy routine:
// (the add/sb can be fused)
           sub     a3, a0, a1
           lbu     a4,0(a1)
           addi    a1,a1,1
           add     a5,a1,a3
           sb      a5,-1(a5)
           bnez    a5,172b04
  \end{lstlisting}
\end{minipage}
\end{center}
  
{\tt 471.omnetpp} is the only SPECInt benchmark for the {\tt
gcc 5.3.0} compiler options used in this study that emitted packed
SIMD operations. These come from the {\tt \_\_strcmp\_sse3}
routine. Meanwhile, it takes 50\% more instructions for RV64G to implement
{\tt strcmp}.

Code \ref{code_omnetpp_strcpy_armv8} compares RV64GC and ARMv8's {\tt strcpy} implementation.

\subsection{473.astar}

{\tt 473.astar} implements a popular AI path-finding routine. 90\% of
the instruction count for {\tt 473.astar} is covered by about 240
RV64G instructions and about 220 \amd\ instructions.  Unsurprisingly,
{\tt 473.astar} is very branch heavy, allowing RV64G to surpass the
other ISAs with the fewest emitted instructions.

\subsection{483.xalancbmk}

The {\tt 483.xalancbmk} benchmarks transformations between XML
documents and other text-based formats.  The reference input generates
a 60 MB text file.

Unadjusted, the {\tt 483.xalancbmk} benchmark is the worst performer
for RISC-V at nearly double the instruction count relative
to \amd. The 34\% most executed instructions are in a spin-loop in
which the simulator waits for the tethered host to service the proxied
I/O request.

\begin{center}
\begin{minipage}{\textwidth}
\begin{lstlisting}[caption=34\% of executed instructions in 483.xalancbmk, label=code_label]
<htif_tty_write>:
loop:
   div   a5, a5, zero
   ld    a5, 24(s0)
   bnez  a5, loop
\end{lstlisting}
\end{minipage}
\end{center}

The divide-by-zero instruction is an interesting quirk of the early
Berkeley silicon RISC-V implementations: it was the lowest energy
instruction that also tied up the pipeline for a number of cycles. A
{\tt Wait For Interrupt} instruction has since been added to RISC-V to
allow processors to sleep while they wait on external agents.
However, WFI is a hint that can be implemented as a NOP (that is how
WFI is handled by the {\tt spike} ISA simulator).

The tethered Host-Target Interface itself is also an artifact of early Berkeley
processors which will eventually be removed entirely from the RISC-V Privileged ISA
Specification.  To prevent the conclusions from being polluted by a simulation
platform artifact, the {\tt htif\_tty\_write} spin loop has been removed from
the data presented in this report.

\begin{center}
\begin{minipage}{\textwidth}
\begin{lstlisting}[caption=the top 9\% of executed user-level instructions in 483.xalancbmk, label=code_label]
<__memcpy>
...
400d3c:  lbu    a5, 0(a1)
400d40:  addi   a4, a4, 1
400d42:  addi   a1, a1, 1
400d44:  sb     a5, -1(a4)
400d48:  bltu   a4, a7,400d3c
...
\end{lstlisting}
\end{minipage}
\end{center}

\end{document}